# Indoor radon ($^{222}$Rn) concentration measurements in Cyprus using high–sensitivity portable detectors

Tassos Anastasiou and Haralabos Tsertos[*]

*Department of Physics, University of Cyprus, Lefkosia, Cyprus*

**Stelios Christofides and George Christodoulides**

*Medical Physics Department, Lefkosia General Hospital, Lefkosia, Cyprus*

(Revised version: 27/01/2003)

**Abstract**

Using high–sensitivity radon ($^{222}$Rn) portable detectors (passive electronic devices of the type RADIM3), the airborne $^{222}$Rn concentration in the interior of various Cypriot buildings and dwellings was measured. For each preselected building and dwelling, a calibrated detector was put into a closed room, and the $^{222}$Rn concentration was registered in sampling intervals of 2 to 4 *hours* for a total counting time of typically 48 *hours*. $^{222}$Rn activity concentrations were found to be in the range of 6.2 to 102.8 *Bq m$^{-3}$*, with an overall arithmetic mean value of (19.3 ± 14.7) *Bq m$^{-3}$*. This value is by a factor of two below the world average (population–weighted) value of 39 *Bq m$^{-3}$*. The total annual

---

[*] Corresponding author. E–mail address: tsertos@ucy.ac.cy, Fax: +357–22339060

Department of Physics, University of Cyprus, P.O.Box 20537, 1678 Lefkosia, Cyprus.

effective dose equivalent to the Cypriot population was calculated to be between 0.16 and 2.6 *mSv* with an overall arithmetic mean value of (0.49 ± 0.37) *mSv*.

## 1. Introduction

$^{222}$Rn is a radioactive gas released during the decay of the Uranium–238 ($^{238}$U) natural series, found in varying amounts in rocks and soils. $^{222}$Rn is odourless, invisible, and without taste, therefore cannot be detected with the human senses.

$^{222}$Rn decays directly into Polonium–218 ($^{218}$Po) which following a chain of β$^-$ and α decays is transformed into Polonium–214 ($^{214}$Po). Both polonium radioisotopes emit alpha particles, which are highly effective in damaging lung tissues. These alpha–emitting $^{222}$Rn decay products have been implicated in a causal relationship with lung cancer in humans. So, breathing high concentrations of $^{222}$Rn can cause lung cancer (Henshaw et al., 1990; Lucie, 1989).

Outdoors, where it is diluted to low concentrations in the air, $^{222}$Rn poses significantly less risk than indoors. In the indoor environment, however, $^{222}$Rn can accumulate to significant levels. The magnitude of $^{222}$Rn concentration indoors depends primarily on the construction of the building and the amount of $^{222}$Rn in the underlying soil. The soil composition under and around a house affects $^{222}$Rn levels and the ease with which $^{222}$Rn migrates into a house. Normal pressure differences between the house and the soil can create a slight under–pressure in the house that can draw $^{222}$Rn gas from the soil into the building. $^{222}$Rn moves more rapidly through permeable soils, such as coarse sand and gravel, than through impermeable soils, such as clays. Fractures in any soil or rock allow $^{222}$Rn to move more quickly. In a very small number of houses, drinking water can also be a significant source of $^{222}$Rn (Otton, 1992).

$^{222}$Rn gas can enter the house from the soil through cracks in concrete floors and walls, floor drains, sump pumps, construction joints, and tiny cracks or pores in hollow–block walls. $^{222}$Rn levels are generally highest in basements and ground floor rooms that are in contact with the soil. Factors such as the design, construction, and ventilation of the house affect the pathways and sources that can draw $^{222}$Rn indoors. Another source of $^{222}$Rn indoors may be the air released by well water during showering and other household activities. Compared to $^{222}$Rn entering the house through soil, $^{222}$Rn entering the house through water will in most cases be a small source of risk (Tanner, 1986).

The higher the uranium level in an area is, the greater the chances are that houses in the area will have high levels of indoor $^{222}$Rn. But some houses in areas with lots of uranium in the soil have low levels of indoor $^{222}$Rn, and other houses on uranium–poor soils have high levels of indoor $^{222}$Rn. Clearly, the amount of $^{222}$Rn in a house is affected by other factors in addition to the presence of uranium in the underlying soil, e.g. the gas permeability of the soil (Reimer, 1988).

From a geological point of view, there should be little radiation threat from $^{222}$Rn in Cypriot buildings and dwellings. The only rock formation that contains significantly elevated levels of uranium in Cyprus is the Plagiogranites of the Troodos mountain Ophiolite (Mukasa and Ludden, 1987).

The Troodos ophiolite consists of basic and ultrabasic pillow lavas, fringed by avdesitic sheeted dykes. The central part consists of basic and ultrabasic plutonic rocks (gabbros, peridotites, dunites and serpentinized harzburgites).

The highly tectonised and fractured conditions of the Troodos mass, a consequence of its uplift, facilitated deep weathering of the rocks, leading to the development of a smooth, mature topography, mantled with a thick cover of a diversity of soils (Robinson and Malpas, 1998). These soils combined with a variety of microclimates produce extensive and renewable forests and a great diversity in the flora. The soil cover of the central area is highly alkaline and provides for very special habitats for certain plants species. The soils on the slopes, lower down, cover sheeted diabase and are neutral. The weathering of the sedimentary rocks (chalks, marls, etc.) in the foothills that fringe Troodos, gave rise to alkaline, calcium rich soils.

Another factor that may increase indoor $^{222}$Rn concentrations in buildings and dwellings in Cyprus is that the Troodos Ophiolite has a large number of deep geological faults. The depth provides a very large surface area over which $^{222}$Rn may emanate (Christofides and Christodoulides, 1993). The cross–sectional area along the length of the faults is much less than the surface area, thus enhancing the concentration of $^{222}$Rn reaching the surface. Many villages are built on top of such faults.

A typical Cypriot house contains strengthened reinforced concrete frame, walls made from bricks, roof from concrete and tiles above the reinforced concrete floor. The walls are covered by plaster. The doors are mainly wooden and the windows are mainly made of glass with exterior cover from the sun. Also, the climatic conditions that prevail in Cyprus allow buildings to be ventilated for long time period, even during the winter. So, the ventilation rate is very high and, therefore, one expects low $^{222}$Rn concentration in the interior of buildings and dwellings. This has been indeed shown by the first indoor $^{222}$Rn measurements in Cypriot homes, which have been carried out eleven years ago (Christofides and Christodoulides, 1993).

A pilot project was recently initiated by the Nuclear Physics Laboratory of the Department of Physics, University of Cyprus, with the objective to systematically register the indoor $^{222}$Rn concentration in Cypriot buildings and dwellings. Other part of the project includes measurement of the terrestrial gamma radiation (Tzortzis et al., 2002) as well as the detection of alpha–emitting radioisotopes by utilizing radioanalytic techniques and high–resolution alpha–spectroscopy. In this paper, the first results from $^{222}$Rn concentration measurements and estimated effective dose rates to the Cyprus population are presented. The present work refers only to the area controlled by the government of the Republic of Cyprus.

## 2. Experimental technique

### 2.1. Radon detection system – RADIM3

For $^{222}$Rn concentration measurements, high–sensitivity modern portable detectors provided with the commercial name "RADIM3" were employed. The Radim3 is a compact and dedicated detector system, designed to directly monitor $^{222}$Rn concentrations, to determine the $^{222}$Rn entry rate and ventilation coefficient (Plch, 2001). The instrument is shown schematically in Figure 1. It incorporates in addition sensors to simultaneously measure the pressure, the temperature and the humidity. Calibration over the whole dynamical range of the instrument is made by the manufacturer (usually by relation to a reference Radim3 instrument) and the accuracy of the calibration is then verified in the State Metrological Institute of the Czech Republic. Verification is achieved by comparing the results of measurement of $^{222}$Rn concentrations provided by the Radim3 instrument and a reference instrument using a secondary ATMOS standard (the ATMOS User Centre is a service of the German Agency, DFD) and then the value of the calibration factor is adjusted, so that the tested instrument yields the same results as ATMOS. The overall uncertainty in the calibration procedure is then equal to ±10%, as the response of the ATMOS reference instrument is known with an error of about ±7% (Plch, 2001).

The maximum $^{222}$Rn concentration that the instrument can measure is 150 *kBq m$^{-3}$* within a time interval of one *hour*, whereas the rather low concentration (activity) of 30 *Bq m$^{-3}$* is determined with a statistical accuracy equal to ±20% (for a counting time of only one *hour*). The instrument response is 0.8 *imp h$^{-1}$* per *Bq m$^{-3}$*. The sampling time can be adjusted from 0.5 to 24 *hours* (Plch, 2001).

The $^{222}$Rn diffuses into the instrument, where a filter collects all airborne $^{222}$Rn decay products. The $^{222}$Rn concentration is determined by measuring the α–activity of $^{218}$Po (the energy of the α–peak appears at ~6 MeV), which is collected by the electric field on the surface of a semiconductor (Si) detector from the chamber of the detection system. The principle of electrostatic precipitation of $^{218}$Po on a surface barrier detector is described in: (Hopke, 1989). The chamber has an optimised spherical geometry half of which is formed by a grid covered with two layers of a dense cloth. This cloth prevents the entrance of airborne $^{222}$Rn decay products and protects the detector against dust and light. Because most of the $^{222}$Rn decay products are positively charged, the vessel is connected to the positive pole of a high–voltage supply and the surface of the Si detector electrode is connected to the negative pole. The positively charged $^{222}$Rn decay products are neutralized by water vapour and other admixtures, and this effect is kept low by employing the highest possible electric field. The instrument uses a stabilized voltage power supply of 2 k$V$ which is controlled by the internal computer (see Fig. 1).

The optimal form of the chamber and the high electric fields applied provide high experimental sensitivity and reduce the influence of humidity, which is less than 15% for a change in the relative humidity of 50%. The results obtained are corrected for humidity. The mean, minimum, and maximum activity concentration values are calculated and the individual results can be read in the display of Radim3 or transferred via a RS232 connector into a PC.

**2.2. Site selection and counting**

Eighty–four buildings and dwellings were selected in 37 different villages and towns in Cyprus as shown in Table 1 according to the geological features and population concentrations. The measurements were carried out over 9 months (beginning of September 2001 to end of May 2002).

Part of the work was also the repetition of certain measurements that had been done before in Cyprus (Christofides and Christodoulides, 1993), so as to have a comparison of the results. In this part of the work we selected regions (buildings and dwellings) that had shown high concentrations of $^{222}$Rn. Another part of the work was to carry out measurements of the $^{222}$Rn concentration in the area of Pano Polemidia, where a number of cases of leukemia were reported.

The procedure followed was to contact by telephone the house owners in the various areas, explain the purpose of the project and following their agreement to arrange an appointment in order to place the detectors in their houses. Drought–free areas in the houses were selected for detector placement such as basements, so that the measurements would show the maximum $^{222}$Rn concentrations. The detectors were placed at a height of approximately 1 *meter*.

Before starting the measurement, the length of the sampling interval (adjustable from 0.5 to 24 *hours*) was entered in the Radim3 instrument, usually 4 *hours*. Therefore, the instrument was typically in operation for 48 *hours*. In this way $^{222}$Rn fluctuations during the measurement could clearly be seen. A typical measurement is shown in figure 2. Prior to the indoor measurements, the outdoor $^{222}$Rn activity concentration (the outdoor $^{222}$Rn background) in Cyprus was measured and showed a typical value of (3.9 ± 0.8) *Bq m$^{-3}$*. This value has not been subtracted from the results obtained in the indoor measurements.

**2.3. Derivation of the indoor radon effective dose rates**

In order to estimate the annual effective doses indoors, one has to take into account the conversion coefficient from absorbed dose in air to effective dose and the indoor occupancy factor. In the UNSCEAR 2000 Report, a value of 9.0 *nSv h$^{-1}$* per *Bq m$^{-3}$* was used for the conversion factor (effective dose received by adults per unit $^{222}$Rn activity per unit of air volume), 0.4 for the equilibrium factor of $^{222}$Rn indoors and 0.8 for the indoor occupancy factor. Hence, the effective dose rate indoors in units of *mSv y$^{-1}$*, $H_E$, is calculated by the following formula:

$$H_E = C_{Rn} \bullet F \bullet T \bullet D \qquad (1)$$

where $C_{Rn}$ is the measured $^{222}$Rn concentration (in *Bq m$^{-3}$*), F is the $^{222}$Rn equilibrium factor indoors (0.4), T is the indoor occupancy time (0.8 × 24 *h* × 365.25 ≅ 7010 *h y$^{-1}$*), and D is the dose conversion factor (9.0 ×10$^{-6}$ *mSv h$^{-1}$ per Bq m$^{-3}$*).

As an arithmetic example, for a measured $^{222}$Rn concentration of 40.0 *Bq m$^{-3}$* the above formula yields an effective dose equivalent to the population which is equal to 1.0 *mSv y$^{-1}$*.

## 3. Results and discussion

Table 1 shows the airborne $^{222}$Rn concentrations in Cypriot buildings and dwellings as well as the arithmetic mean (A.M.), the standard deviation (S.D.), the minimum (Min) and the maximum (Max) $^{222}$Rn concentration. As shown in Table 1, $^{222}$Rn activity concentrations ranged from 6.2 to 102.8 *Bq m$^{-3}$*, with an overall arithmetic mean value of (19.3 ± 14.7) *Bq m$^{-3}$*. Out of the 37 areas measured in this study, a dwelling at Agia Zoni in the district of Lemesos appears to have the highest $^{222}$Rn concentration (102.8 ± 10.1) *Bq m$^{-3}$*. The second highest $^{222}$Rn concentration appears in a dwelling at Egkomi in the district of Lefkosia and is (68.6 ± 5.7) *Bq m$^{-3}$*. These two dwellings were investigated further to find out if there is a reason for their relatively high concentration values. The detectors were placed in the basements of the dwellings. The basements were closed for the entire duration of the measuring time period (48 *hours*). The observed high values may be due to a very low air change rate compared to the normal Cypriot living conditions. On the other hand, the area of Latsia in the district of Lefkosia appears to have the lowest $^{222}$Rn concentration (6.2 ± 0.8) *Bq m$^{-3}$*.

In order to obtain a more general and representative overview, the measurements were classified in 11 main regions of Cyprus, shown in Table 2. In this way, the results from various measurements belonging to the same region were averaged. As can be seen in Table 2, the highest mean $^{222}$Rn concentration appears in the Lemesos region and is (29.6 ± 10.7) *Bq m$^{-3}$*, whereas the lowest mean $^{222}$Rn concentration appears in the centre of Larnaka region and is (8.6 ± 1.4) *Bq m$^{-3}$*.

Figure 3 illustrates the indoor $^{222}$Rn concentration for all the 84 buildings and dwellings studied together with their statistical errors, which are in the range of 10–20%. The horizontal bold line shows the arithmetic mean value of 19.3 *Bq m$^{-3}$*. Most of the individual measurements are around the arithmetic mean value.

Finally, the measured $^{222}$Rn activity concentration was converted into an indoor effective dose equivalent to the Cypriot population using equation (1), for all the 84 buildings and dwellings studied. The results are plotted in figure 4 together with their statistical errors. The horizontal bold line shows the arithmetic mean value of 0.49 *mSv y$^{-1}$*.

In summary, the present systematic investigations on $^{222}$Rn activity concentration in all the representative villages and towns clearly demonstrate that the $^{222}$Rn concentration is very low in Cyprus. The overall mean value derived (19.3 *Bq m$^{-3}$)* is by a factor of two lower than the reported world average value of 39 *Bq m$^{-3}$* (UNSCEAR 2000 Report). The results are in the general trend consistent with previous observations (Christofides and Christodoulides, 1993). They are also consistent with recent investigations on terrestrial gamma–radiation measurements in all the predominant rock formations appearing in Cyprus, which have revealed low activity and elemental concentrations of uranium and thorium radioisotopes (Tzortzis et al., 2002).

## 4. Conclusions

Airborne $^{222}$Rn concentration in dwellings and buildings in most countries is the major contributor to the annual effective dose equivalent to the population (Nero et al, 1986, Clarke and Southwood, 1989). From the present systematic measurements, an average indoor annual effective dose equivalent to the Cypriot population of (0.49 ± 0.37) *mSv* is calculated, if the mean value of (19.3 ± 14.7) *Bq m$^{-3}$* and a $^{222}$Rn equilibrium factor of 0.4 are used. The extracted value of the annual effective dose equivalent to the Cypriot population is almost 2 to 3 times less than that quoted in the literature for other countries worldwide. Note that the world average $^{222}$Rn concentration value is reported to be 39 *Bq m$^{-3}$* (UNSCEAR 2000 Report), which corresponds to an annual effective dose to the population of about 1 *mSv*.

The indoor $^{222}$Rn concentration results show a correlation with the geological formations on which the buildings and dwellings were built. The highest $^{222}$Rn concentrations were in buildings and dwellings built on chalk formations.

Deductively, the results show that the threat from $^{222}$Rn in Cypriot buildings and dwellings is negligible and much lower than the internationally acceptable levels for safety, which certain countries have recommended. In particular for the region of Pano Polemidia, the leukemia cases observed there cannot have their origin in high $^{222}$Rn concentrations.

## Acknowledgements

This work is conducted with financial support from the Cyprus Research Promotion Foundation (Grant No. 45/2001) and partially by the University of Cyprus.

**TABLE CAPTIONS**

**Table 1.** Airborne $^{222}$Rn concentration in the 84 buildings and dwellings selected in 37 different villages and towns of Cyprus indicated.

| District | A/A | Area | Samples | $^{222}$Rn Concentration [$Bq\ m^{-3}$] | | | |
|---|---|---|---|---|---|---|---|
| | | | | A.M.[†] | S.D. | Min | Max |
| Lefkosia | 1 | Agios Andreas | 2 | 11.5 | 2.1 | 7.0 | 14.3 |
| | 2 | Agios Dometios | 2 | 16.7 | 3.4 | 4.7 | 32.4 |
| | 3 | Aglangia | 10 | 18.2 | 14.5 | 2.7 | 84.1 |
| | 4 | Akropoli | 2 | 10.5 | 9.6 | 1.1 | 38.9 |
| | 5 | Anthoupoli | 2 | 22.6 | 3.4 | 6.7 | 32.6 |
| | 6 | Geri | 2 | 20.0 | 2.8 | 9.3 | 37.9 |
| | 7 | Egkomi | 1 | 68.6 | 5.7 | 31.5 | 111.0 |
| | 8 | Kaimakli | 15 | 21.5 | 12.0 | 3.1 | 77.3 |
| | 9 | Lakatamia | 2 | 11.9 | 1.6 | 5.8 | 15.8 |
| | 10 | Latsia | 2 | 6.2 | 1.1 | 3.5 | 10.8 |
| | 11 | Makedonitissa | 2 | 36.4 | 12.2 | 30.1 | 43.1 |
| | 12 | Palouriotissa | 2 | 20.0 | 4.0 | 8.9 | 79.5 |
| | 13 | Lefkosia centre | 3 | 19.4 | 19.6 | 3.4 | 55.7 |
| | 14 | Strovolos | 5 | 22.4 | 7.6 | 1.7 | 58.0 |
| | 15 | Tseri | 4 | 7.5 | 2.8 | 2.7 | 18.3 |
| Larnaka | 16 | Xylofagou | 1 | 16.6 | 4.1 | 10.2 | 25.1 |
| | 17 | Perivolia | 1 | 15.8 | 4.0 | 9.8 | 21.3 |
| | 18 | Larnaka centre | 2 | 8.6 | 2.0 | 6.3 | 12.7 |
| | 19 | Pyla | 1 | 19.6 | 4.4 | 15.7 | 24.8 |

[†] "A.M." means the arithmetic mean, "S.D." the standard deviation due to the various measurements, "Min" the minimum and "Max" the maximum radon concentration.

**Table 1.** (continued)

| District | A/A | Area | Samples | $^{222}$Rn Concentration [$Bq\ m^{-3}$] | | | |
|---|---|---|---|---|---|---|---|
| | | | | A.M. | S.D. | Min | Max |
| Lemesos | 20 | Agia Zoni | 1 | 102.8 | 10.1 | 34.8 | 183.5 |
| | 21 | Agia Fyla | 1 | 26.6 | 5.2 | 3.4 | 50.0 |
| | 22 | Agios Athanasios | 2 | 22.8 | 3.1 | 5.6 | 38.1 |
| | 23 | Agios Mamas | 1 | 9.3 | 3.0 | 5.3 | 12.9 |
| | 24 | Akrounta | 1 | 12.9 | 3.6 | 8.0 | 17.8 |
| | 25 | Germasogia | 1 | 21.7 | 4.7 | 4.4 | 50.4 |
| | 26 | Limnatis | 1 | 19.4 | 4.4 | 10.0 | 27.9 |
| | 27 | Mouttagiaka | 1 | 13.4 | 3.7 | 1.7 | 25.6 |
| | 28 | Pano Polemidia | 4 | 20.2 | 12.2 | 0.9 | 48.2 |
| | 29 | Parekklisia | 1 | 20.6 | 4.5 | 3.8 | 36.2 |
| | 30 | Lemesos centre | 1 | 23.5 | 4.8 | 6.3 | 93.0 |
| | 31 | Potamos Germasogias | 1 | 6.2 | 2.5 | 1.7 | 11.0 |
| Ammochostos | 32 | Agia Napa | 1 | 11.2 | 3.3 | 5.5 | 16.1 |
| | 33 | Avgorou | 2 | 14.8 | 0.3 | 7.5 | 19.3 |
| | 34 | Frenaros | 1 | 6.9 | 2.6 | 2.0 | 14.3 |
| Pafos | 35 | Pano Arodes | 1 | 12.8 | 3.6 | 2.8 | 23.0 |
| | 36 | Pafos centre | 1 | 15.7 | 4.0 | 2.9 | 66.9 |
| | 37 | Filousa | 1 | 8.6 | 2.9 | 1.1 | 44.5 |

**A.M. ± S.D. = (19.3 ± 14.7) $Bq\ m^{-3}$**

Table 2. Mean $^{222}$Rn concentration in 11 main regions of Cyprus indicated.

| A/A | Region | Samples | $^{222}$Rn Concentration [$Bq\ m^{-3}$] | | | |
|---|---|---|---|---|---|---|
| | | | A.M. | S.D. | Min | Max |
| 1 | Ammochostos | 4 | 11.9 | 3.8 | 2.0 | 19.3 |
| 2 | Larnaka | 3 | 17.3 | 2.1 | 9.8 | 25.1 |
| 3 | Lemesos | 8 | 29.6 | 30.3 | 1.7 | 183.5 |
| 4 | Lefkosia | 53 | 19.4 | 13.1 | 1.1 | 111.0 |
| 5 | Lemesos forest area | 3 | 13.9 | 5.0 | 5.3 | 27.9 |
| 6 | Pano Polemidia | 4 | 20.2 | 12.2 | 0.9 | 48.2 |
| 7 | Pafos | 2 | 10.7 | 3.0 | 1.1 | 44.5 |
| 8 | Larnaka centre | 2 | 8.6 | 2.0 | 6.3 | 12.7 |
| 9 | Lemesos centre | 1 | 23.5 | 4.8 | 6.3 | 93.0 |
| 10 | Lefkosia centre | 3 | 19.4 | 19.6 | 3.4 | 55.7 |
| 11 | Pafos centre | 1 | 15.7 | 4.0 | 2.9 | 66.9 |

**FIGURE CAPTIONS**

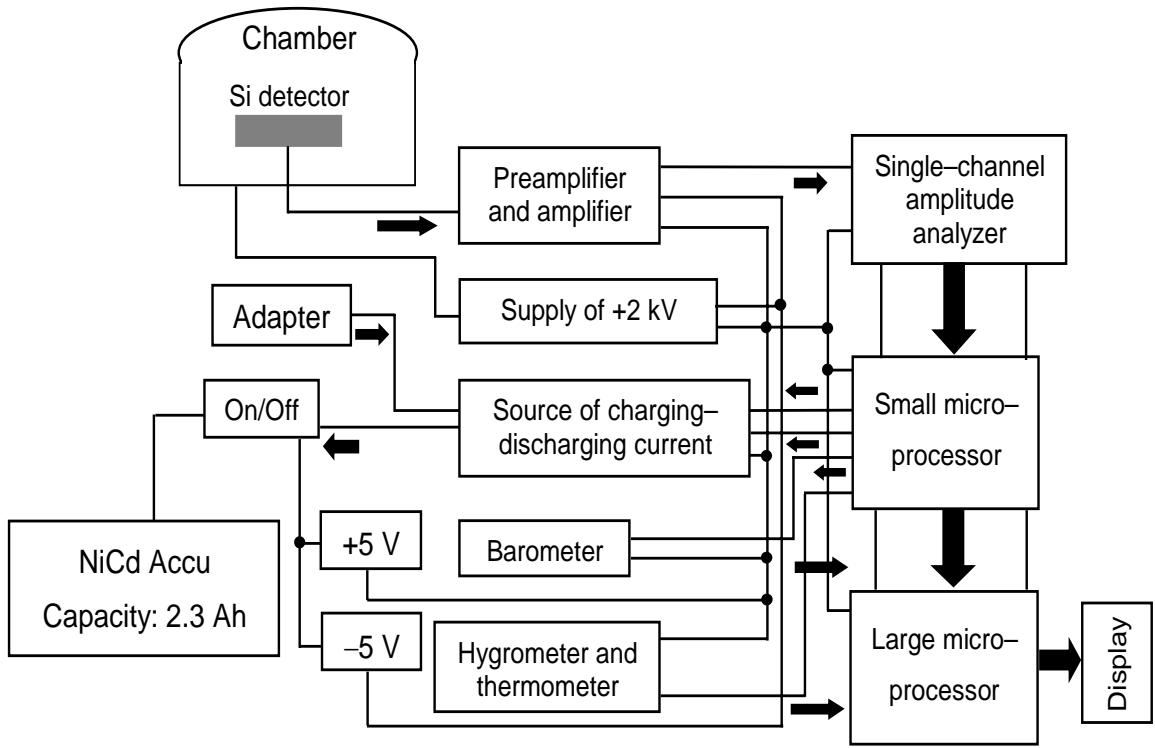

**Figure 1.** Block diagram of the $^{222}$Rn detection system RADIM3.

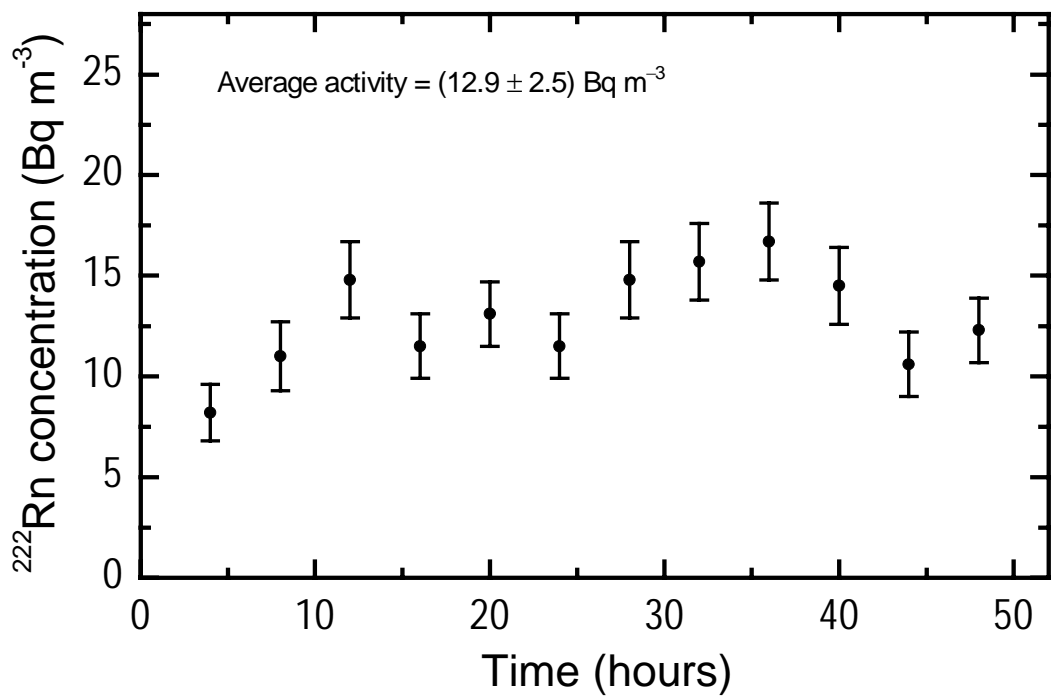

**Figure 2.** Typical measurement of indoor $^{222}$Rn concentration as a function of time for a dwelling in the Kaimakli area. Sampling time was 4 *hours* and started at 16:15. The error bars are solely of statistical origin. The observed diurnal variation (maximum activity during midnight and minimum in the afternoon) is due to a corresponding change in the air ventilation rate.

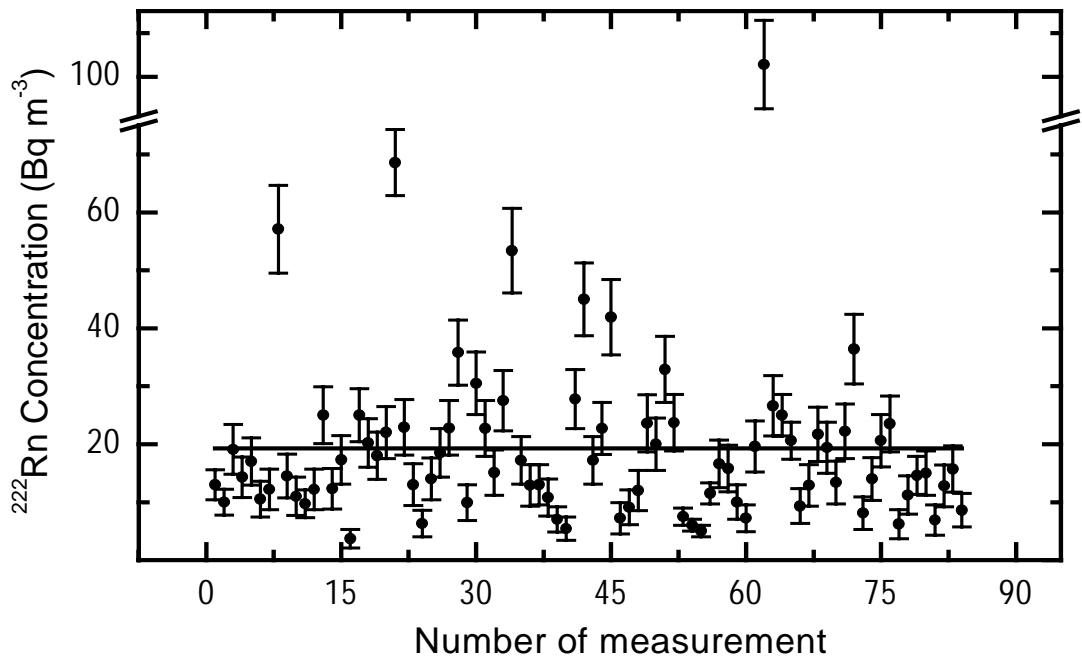

**Figure 3.** Indoor $^{222}$Rn concentration for all the 84 buildings and dwellings studied. The error bars are due to counting statistics. The horizontal bold line represents the overall arithmetic mean value of 19.3 *Bq m$^{-3}$*. The numbers in the abscissa correspond to the villages and towns given in Table 1 in consecutive manner.

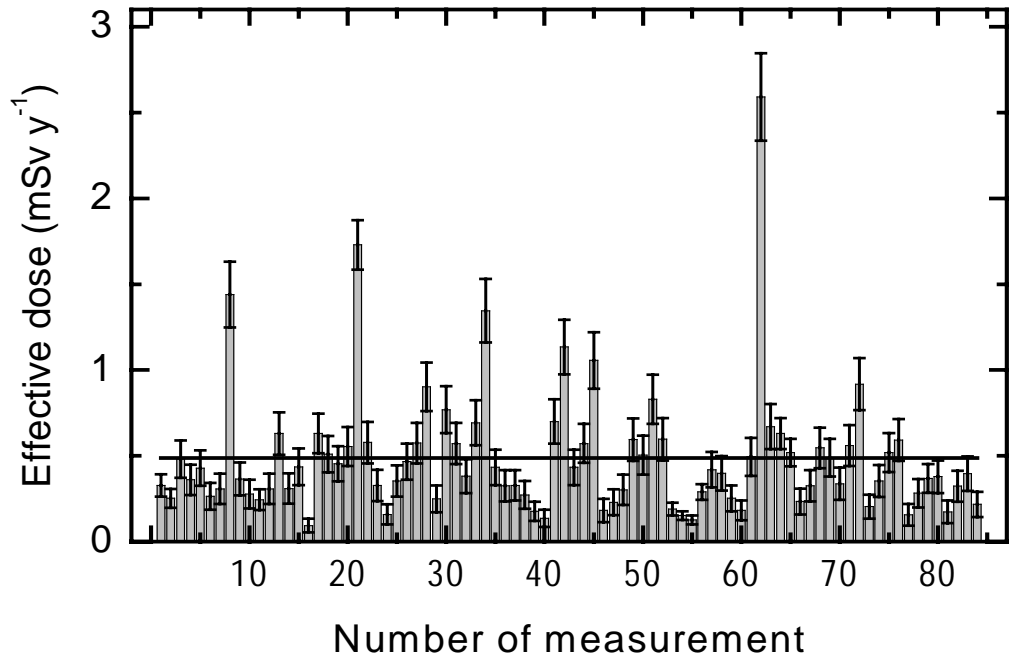

**Figure 4.** Effective dose equivalent to the Cypriot population calculated from indoor $^{222}$Rn concentration in 84 buildings and dwellings studied (see Table 1). The error bars are due to counting statistics. The horizontal bold line represents the overall arithmetic mean value of 0.49 *mSv*.